\documentclass{midl} % Include author names
\usepackage{mwe} % to get dummy images

\jmlrproceedings{MIDL}{Medical Imaging with Deep Learning}
\jmlrpages{}
\jmlryear{2023}

\jmlrworkshop{Short Paper -- MIDL 2023 submission}
\jmlrvolume{-- Under Review}
\editors{Under Review for MIDL 2023}

\title[X-ray Recognition]{X-ray Recognition: Patient identification from X-rays using a contrastive objective}

 \midlauthor{\Name{Hao Liang} \Email{hl106@rice.edu}\\
  \Name{Kevin Ni} \Email{kn44@rice.edu}\\
  \Name{Guha Balakrishnan} \Email{guha@rice.edu}\\
  \addr Department of Electrical and Computer Engineering, Rice University, USA}

\begin{document}

\maketitle

\begin{abstract}
Recent research demonstrates that deep learning models are capable of precisely extracting bio-information (e.g. race, gender and age) from patients' Chest X-Rays (CXRs). In this paper, we further show that deep learning models are also surprisingly accurate at \emph{recognition}, i.e., distinguishing CXRs belonging to the same patient from those belonging to different patients. These findings suggest potential privacy considerations that the medical imaging community should consider with the proliferation of large public CXR databases.
\end{abstract}


\section{Introduction}

\begin{figure}[t!]
    \centering
    
    {\label{fig:1}\includegraphics[width=\textwidth]{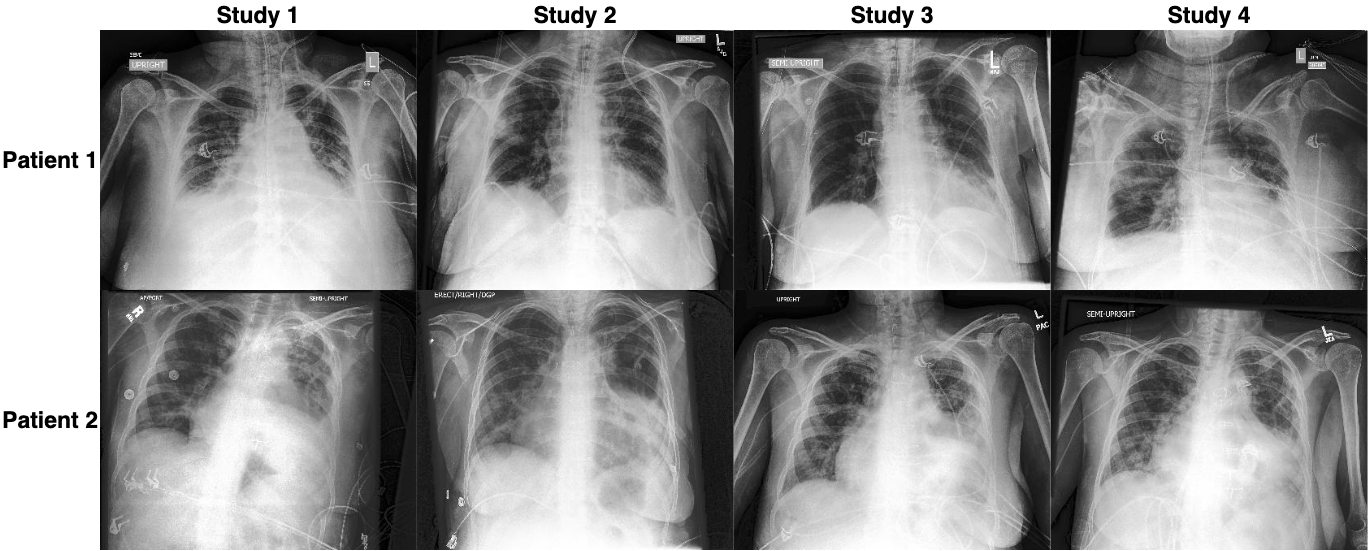}}
    % \vspace{-1em}
    \caption{\textbf{Example CXRs from Chexpert.} We show four CXRs for two random patients taken over multiple visits. Various changes, related to health condition and other factors make it hard to visually determine whether a pair of CXRs belong to the same person or not.}
    \label{fig:example_chex}
% \vspace{-1.5em}
\end{figure}

Identifying patient bio-information such as age, race, and gender from Chest X-ray (CXR) scans has attracted recent interest in the context of exploring demographic-specific biases and confounders~\cite{gichoya2022ai,karargyris2019age,duffy2022confounders}. The ``Reading Race'' study in particular~\cite{gichoya2022ai} demonstrated that a patient's race may be predicted with over $0.99$ AUROC from CXR datasets, raising questions on what visual cues may be unknowingly exploited by downstream algorithms. 
%In these studies, researchers build deep learning models using self-reported race/age/gender labels from the patients, and demonstrate that these models achieve high predictive accuracy. Some of their findings are surprising, such as 
Inspired by these studies, we introduce the natural follow-up challenge of \emph{X-ray Recognition}: identifying whether a pair of X-rays belong to the same person or not. Interestingly, this task is challenging even for a human to perform due to the dramatic variations in CXRs across patient visits, as demonstrated in Fig.~\ref{fig:example_chex}. Yet in this study, we demonstrate that a deep learning model can achieve impressive accuracies on this task on public CXR datasets. 

CXR datasets typically contain multiple images from the same patient across patient visits. For example, in the popular ChexPert~\cite{irvin2019chexpert} and NIH CXR datasets~\cite{jaeger2014two}, $48.7\%$ and $43.2\%$ percent of patients have two or more CXRs. Using this fact, we design our X-ray recognition model based on the typical design of modern face recognition models~\cite{schroff2015facenet,liu2017sphereface}. In particular, we employ a \emph{contrastive, self-supervised} training procedure, which learns a deep feature space by maximizing the distance between CXRs from different patients, while minimizing the distance CXRs belonging to the same patient.
%We hypothesize that a successful X-ray recognition model must learn a complex set of visual features to properly match image pairs with such a high degree of variability. During training, we propose an X-ray recognition convolutional neural network (CNN) based on the design of state-of-the-art face recognition models. 
 Results on public CXR datasets demonstrate that the proposed model achieves surprisingly high accuracy ($0.99$ AUROC) on ChexPert and that its learned features are useful for other bio-information prediction tasks. Given that CXRs are one of the most common examinations in the clinic and that there is growing interest in collecting large public CXR datasets for AI research, these findings suggest potential privacy concerns that the community should consider.

%with the proliferation of large CXR datasets to the public. 

%An X-ray recognition model may serve as a useful tool for identifying patients, as well as downstream tasks like transfer learning and image registration, etc. However, it also raises the concern of patient privacy leakage since CXR has been the most popular examination during hospital visit.

\section{Method}
\label{sec:method}
% \begin{figure}[t!]
%     \centering
    
%     {\label{fig:2}\includegraphics[width=1\textwidth]{imgs/frame.drawio.png}}
%     % \vspace{-1em}
%     \caption{\textbf{CXR recognition \& triplet loss}. (a)CXR recognition: given a pair of CXRs, the task of the model is to predict whether they are from the same person or not. (b) triplet loss: during training, we deploy a triplet loss to  maximize the distance between the features that are from CXRs of different patients, while minimize the distance between the features that are from CXRs of the same patient.}
%     \label{fig:frame}
% % \vspace{-1.5em}
% \end{figure}

% \begin{comment}
% \begin{figure}[hbtp]
%     \centering
    
%     {\label{fig:1}\includegraphics[width=\textwidth]{imgs/x_ray_example_nih.drawio.png}}
%     % \vspace{-1em}
%     \caption{\textbf{Example CXRs from NIH.} We randomly choose 2 patients' CXRs (4 for each) from NIH. The CXRs of the same patient were usually taken from different time, possibly with different health condition (i.e. recovered from/caught a specific disease), which makes it hard to distinguish whether they are from the same person or not from human's perspective. Also, CXRs from different patients can sometimes look close, also makes the task hard to humans. [defer to Supp.]}
%     \label{fig:res1}
% % \vspace{-1.5em}
% \end{figure}
% \end{comment}

We design our model and training objective based on recent face recognition works~\cite{schroff2015facenet,liu2017sphereface}. Let $f(\cdot)$ denote our recognition model, which takes a CXR image as input, and returns an embedding vector $y \in R^{d}$ that intuitively represents the ``identity'' information within the CXR. In our experiments, we use DenseNet-121 \cite{huang2017densely} to implement $f$, a powerful model architecture used in various computer vision problems, and replace the final classifier layer with a fully-connected layer that output a $d=512$-dimension vector. 

During training, we construct {\it positive} and {\it negative} image pairs from the training dataset. A positive pair consists of two CXRs from the same patient and a negative pair consists of two CXRs from different patients. We use a \textit{triplet loss} to train $f(\cdot)$: 
\begin{equation}
L(x^a, x^p, x^n) = \max (||f(x^a) - f(x^p)||^2_2 - ||f(x^a) - f(x^n)||^2_2 + \alpha, 0),
\end{equation}
where $x^a$ is known as an ``anchor'' image, $x^p$ is a CXR of the same patient (positive), and $x^n$ is a CXR from another patient (negative). The loss encourages the anchor and positive CXRs to lie close together in the embedding feature space, while simultaneously maximizing the distance between the anchor and negative examples. $\alpha$ is a hyperparameter quantifying a desired margin between positive and negative pairs.

At inference time, we give a pair of CXRs ($x_A, x_B$) to the model $f$, which returns a similarity score $S = || f(x_A) - f(x_B)||_2^2$. If $S \geq t$, where $t$ is a threshold chosen during the validation stage, the pair is predicted to belong to the same patient.
% Acknowledgments---Will not appear in anonymized version
% \midlacknowledgments{We thank a bunch of people.}

\begin{figure}[t!]
    \centering
    
    {\label{fig:1}\includegraphics[width=1\textwidth]{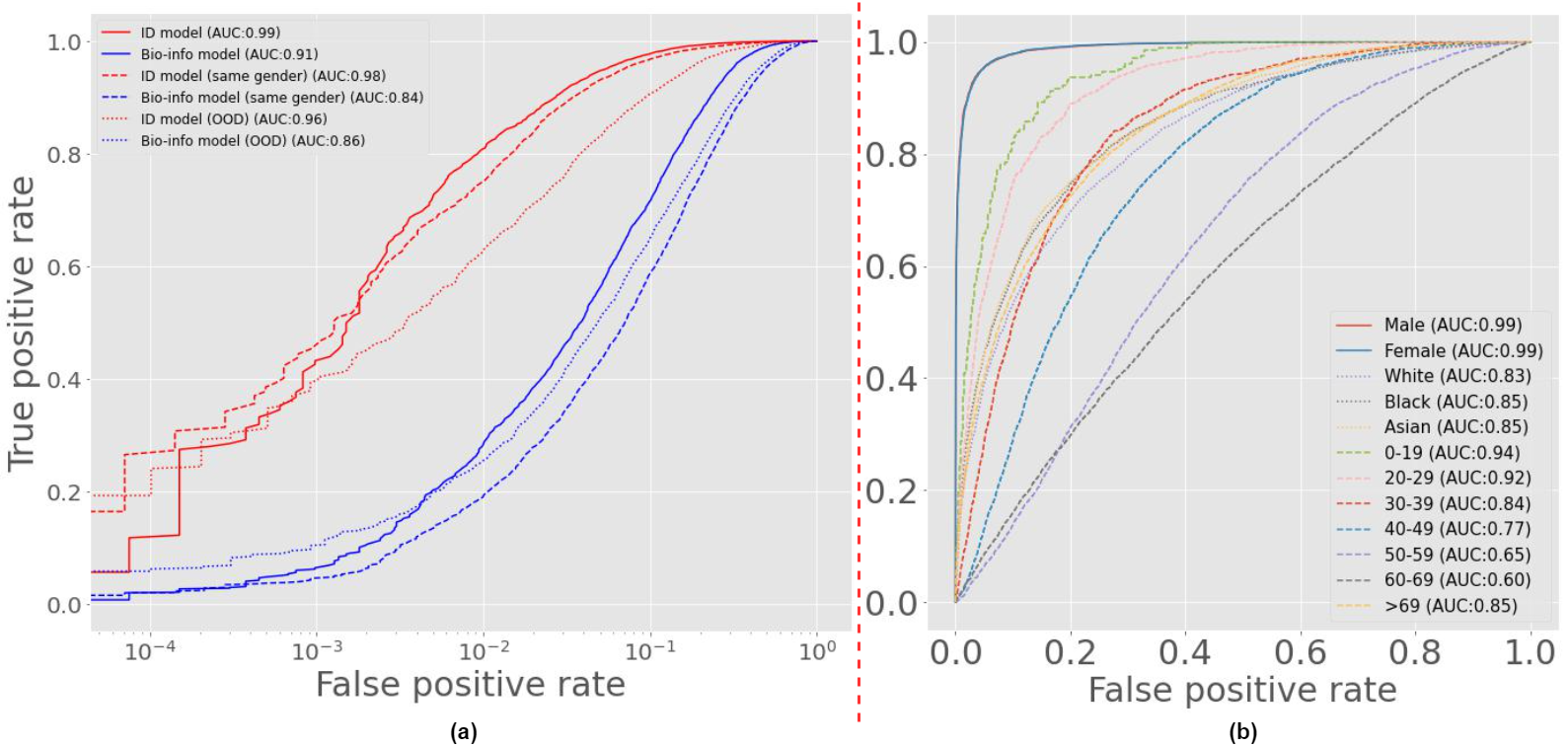}}
    % \vspace{-1em}
    \caption{\textbf{(a) True Positive Rate(TPR) vs. False Positive Rate(FPR) and AUROC scores for recognition models on different test dataset settings.} We considered two models: 1. The proposed X-Ray recognition model (red), 2. DenseNet-121 trained for bio-information classification (baseline, black). We trained the second model for its respective tasks, and use its feature .. We consider the following test set settings: 1. randomly selected positive/negative pairs (solid line), 2. negative pairs with the same gender (dashed line), and 3. OOD test set from NIH dataset (dotted line).  Results show that recognition model is able to distinguish IDs under all settings with a much higher accuracy. \textbf{(b) Results of transfer learning on bio-information prediction.} Results show that one can make prediction with high accuracy only based on information captured by the recognition model.}
    \label{fig:perf}
% \vspace{-1.5em}
\end{figure}

\section{Experiments and Conclusion}
We evaluate our model on ChexPert~\cite{irvin2019chexpert}. We trained our model on ChexPert's training dataset, and considered three test dataset settings and two other baseline models. Please refer to the caption of Fig.~\ref{fig:perf}(a) for experimental details and results.

We also demonstrate the ability of the recognition features to transfer to other bio-information prediction tasks. To do so, we freeze the weights of the whole model, add one trainable fully-connected linear layer, and optimize this final layer for each prediction task. Results are in Fig.~\ref{fig:perf}(b).
 % : 1. randomly selected positive/negative pairs, in this setting, we construct all the pairs randomly from Chexpert, with {\it identity} as the only interested attribute; 2. negative pairs with the same gender, this is because previous researches show that deep learning models are able to capture the information of sex from CXRs easily, which might be used as a key feature in the ID recognition procedure; 3. OOD test set, to confirm the ability of generalization of our model, we also tested it on the NIH test dataset. On the other hand, to show that the model is able to non-trivially distinguish patients' ID, we also compare it with two other models: 1. a DenseNet-121 trained for multi-disease classification (14 classes); 2. a DenseNet-121 trained for bio-information(age\&race\&gender) classification. These two models are trained to have a comparable performance with the popular public models. We removed the last fully-connected layer of them, and treat the output of the models as the features captured. 

Our results show that deep learning models are able to achieve surprisingly high accuracy on X-ray Recognition. The features also clearly contain a rich set of information of bio-information like gender, race, and age based on our transfer learning experiments. Based on these results, we believe the issue of privacy violations in large CXR datasets could be a potential issue for the medical imaging community to consider in the future.

\clearpage 
\bibliography{midl-samplebibliography}

\end{document}